\documentclass{article}
\usepackage[utf8]{inputenc}
\usepackage{graphicx}
\usepackage{bm,revsymb,fancybox,ascmac}
\usepackage{amsmath}						
\usepackage{amssymb}					
\usepackage{amsfonts}	
\usepackage{authblk}
\usepackage{braket} 
\usepackage[margin=20truemm]{geometry}

\begin{document}

\title{A large-scale single-mode array laser based on a topological edge mode}
\date{}	


\author[1]{Natsuko Ishida}
\author[2]{Yasutomo Ota}
\author[3]{Wenbo Lin} 
\author[4]{Tim Byrnes}
\author[5]{Yasuhiko Arakawa}
\author[6]{Satoshi Iwamoto} 

\affil[1]{\protect\raggedright 
Research Center for Advanced Science and Technology, 
The University of Tokyo, 4-6-1 Komaba, Meguro, 
Tokyo 153-8505, Japan, 
e-mail: n-ishida@iis.u-tokyo.ac.jp}
\affil[2]{\protect\raggedright 
Department of Applied Physics and Physico-Informatics,
Keio University, 3-14-1 Hiyoshi, Kohoku-ku, Yokohama, Kanagawa 223-8522, Japan; 
Institute for Nano Quantum Information Electronics, 
The University of Tokyo, 4-6-1 Komaba, Meguro,
Tokyo 153-8505, Japan, e-mail: ota@appi.keio.ac.jp}
\affil[3]{\protect\raggedright Research Center for Advanced Science and Technology, 
The University of Tokyo, 4-6-1 Komaba, Meguro, 
Tokyo 153-8505, Japan; 
Institute of Industrial Science, 
The University of Tokyo, 4-6-1 Komaba, Meguro, 
Tokyo 153-8505, Japan}
\affil[4]{\protect\raggedright New York University Shanghai, 1555 Century Ave, Pudong, Shanghai 200122, China;
State Key Laboratory of Precision Spectroscopy, School of Physical and Material Sciences, East China Normal University, Shanghai 200062, China; 
NYU-ECNU Institute of Physics at NYU Shanghai, 3663 Zhongshan Road North, Shanghai 200062, China; 
National Institute of Informatics, 2-1-2 Hitotsubashi, Chiyoda-ku, Tokyo 101-8430, Japan; 
Department of Physics, New York University, New York, NY 10003, USA}
\affil[5]{\protect\raggedright Institute for Nano Quantum Information Electronics, 
The University of Tokyo, 4-6-1 Komaba, Meguro,
Tokyo 153-8505, Japan}
\affil[6]{\protect\raggedright Research Center for Advanced Science and Technology, 
The University of Tokyo, 4-6-1 Komaba, Meguro, 
Tokyo 153-8505, Japan; 
Institute for Nano Quantum Information Electronics, 
The University of Tokyo, 4-6-1 Komaba, Meguro,
Tokyo 153-8505, Japan;
Institute of Industrial Science, 
The University of Tokyo, 4-6-1 Komaba, Meguro, 
Tokyo 153-8505, Japan}

	
\maketitle

\abstract{Topological lasers have been intensively investigated as a strong candidate for robust single-mode lasers. A typical topological laser employs a single-mode topological edge state, which appears deterministically in a designed topological bandgap and exhibits robustness to disorder. These properties seem to be highly attractive in pursuit of high power lasers capable of single mode operation.
In this paper, we theoretically analyze a large-scale single-mode laser based on a topological edge state. We consider a sizable array laser consisting of a few hundreds of site resonators, which support a single topological edge mode broadly distributed among the resonators. 
We build a basic model describing the laser using the tight binding approximation and evaluate the stability of single mode lasing based on the threshold gain difference $\Delta\alpha$ between the first-lasing edge mode and the second-lasing competing bulk mode.
Our calculations demonstrate that stronger couplings between the cavities and lower losses are advantageous for achieving stable operation of the device. 
When assuming an average coupling of 100 cm$^{-1}$ between site resonators and other realistic parameters, the threshold gain difference $\Delta\alpha$ can reach about 2 cm$^{-1}$, which would be sufficient for stable single mode lasing using a conventional semiconductor laser architecture. 
We also consider the effects of possible disorders and long-range interactions to assess the robustness of the laser under non-ideal situations. 
These results lay the groundwork for developing single-mode high-power topological lasers.}

\section{Introduction}

High power semiconductor lasers have been of great interest to the industrial market for their wide applications, prompting enormous efforts in improving their performance. 
A straightforward approach for increasing laser output power is to widen the emitting area, as adopted in tapered\cite{tapered1996}, broad-area\cite{Crump2013} and array lasers\cite{Fan2005}. 
However, wider emitting areas in general result in multi-mode lasing and thereby in the degradation of laser beam quality. 
To overcome this issue, various techniques have been examined to minimize the effects of unwanted lateral guided modes over the last few decades\cite{Ackley1986,Glova2003,Weyers2014,Erbert2012,Medina2021}.
Singlemodeness can often be improved by a delicate cavity design that cleverly takes advantages of the difference between the spatial mode profiles of the target and other undesired modes\cite{Sarangan1996,Sumpf2009,Zheng2019,Forbes2018,Marsh2005,Harder2006,Erbert2008,Christodoulides2012,Feng2014,Miri2019,SUSY2019}.
A remarkable example reported recently is based on a two-dimensional photonic crystal band-edge resonator with 10W-class output from a single optical mode\cite{Noda2019}. 
However, the designs of these structures tend to be highly delicate and sometimes significantly complicate the fabrication process of the device. 
Such complexity in design may motivate to find a simpler scheme that enables high power single mode semiconductor lasers.

A potential approach in this direction is that of topological lasers, which leverage topological photonics for designing lasing optical modes
\cite{Bahari2017,Amo2017,Khajavikhan2018,Segev2018,Parto2018,Zhao2018,polariton2018,Ota2018,PhC2019,bulklaser2020,Wang2020,corner2020-1,corner2020-2,RenMinMa2020,Kante2020}.
Topological photonics offers a novel route for designing
optical modes with distinctive properties compared to conventional
approaches\cite{Soljacic2014,ShvetsReview2017,Ozawa2019,Ota2020,Nonlinear2020,IwamotoReview2021}.
A typical topological laser consists of a single topological edge mode that deterministically appears in a topological bandgap as a result of a topological mechanism called the bulk-edge correspondence\cite{Jackiw1976,Hatsugai1993a,Hatsugai1993b}.
The topological edge modes are known to behave robustly against certain perturbations due to topological protection, which is suitable for developing robust single mode lasers. 
Topological ring lasers have been demonstrated using one-dimensional topological edge states propagating at the exterior of the bulk emulating quantum Hall\cite{Bahari2017}, quantum spin Hall \cite{Khajavikhan2018,RenMinMa2020} and quantum valley Hall systems \cite{Kante2020,Wang2020}. 
Single mode lasing devices have been demonstrated in these systems and the possibility of realizing robust single mode lasers with high slope efficiencies has been discussed. 
More recently, an electrically pumped topological laser has also been reported at mid-infrared wavelengths\cite{Wang2020}.
Surface-emitting lasers utilizing Dirac cones or those with mass vortices also have been discussed as another candidate for a large-area laser\cite{DiracLaser2014,DiracVortex2020}.

Topological lasers based on zero dimensional edge states are another topic that has gathered interest recently. 
Localized topological modes in arrays of resonators, such as micropillars\cite{Amo2017} and microring cavities\cite{Parto2018,Zhao2018}, have been combined with semiconductor gain to demonstrate lasing. 
Topological nanolasers have also been studied using topological photonic crystals supporting zero dimensional interface states \cite{Ota2018,PhC2019} and corner states as higher-order topological states as well\cite{OtaCorner2019,Cornerlasing2020}.
So far, most of the works employing zero dimensional topological states aimed to investigate the lasing properties of tightly localized topological edge modes or to explore the physics of non-Hermitian topology therein\cite{Weimann2017,Parto2018,Zhao2018,Takata2017,Takata2018,Henning2015,Henning2018}. 
As such, there have been limited discussions for the application of topological edge modes for high power lasers by significantly expanding the mode profile in space.

In this paper, we theoretically investigate a large-scale single-mode topological laser. We consider a sizable array laser that supports a single zero-dimensional topological edge state distributed over a few hundreds of site resonators. We formulate this model based on the tight binding approximation.
Akin to a conventional analysis of semiconductor lasers\cite{DelAlpha1990,Henmi1988,Noda2019}, we analyze the stability of single mode operation by evaluating the threshold gain difference between the first topological mode lasing and the second bulk mode lasing. 
We find that the stability of the single mode lasing increases with a stronger coupling of the site resonators and reducing optical loss in them. 
Furthermore, we study the robustness of the single mode lasing under the presence of imperfections. From the discussion, we deduce a possible direction of the device design for robust single mode lasing with high output power. 
We believe our results pave a new path towards single mode high power lasers based on topological photonics.


\section{Characteristics of an ideal topological edge-mode laser}
\label{sec:ideal}
\subsection{Theoretical model}
\begin{figure}[h!]
\centering\includegraphics[width=14cm]{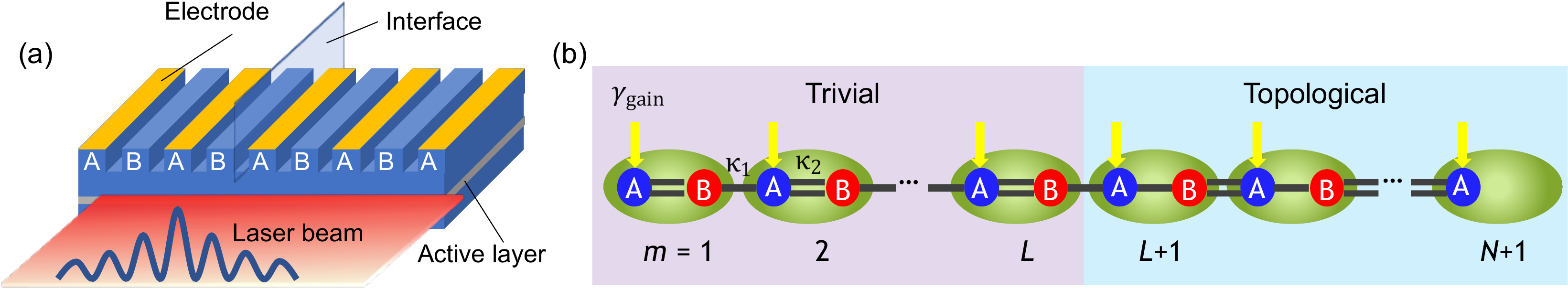}
\caption{(a)Schematic of the investigated topological laser structure.
(b)Tight-binding model of the laser structure. An edge mode
deterministically emerges at the interface of the two topologically distinct photonic lattices. $N$ and $L$ correspond to the total number of dimers and the number specifying the topological phase boundary, respectively. Note that there exists a single auxiliary site at the end
of the topological chain. $\kappa_1$ (single line) and $\kappa_2$ (double lines) indicate weak and strong coupling strengths between neighboring sites, respectively. $\gamma_{\rm{gain}}$ expresses gain supplied on A-site. The application of this tight binding model to the system described in (a) would be valid as long as the longitudinal cavity modes behave independently or when assuming arrays of cavities supporting only single
longitudinal mode such as $\lambda/4$-shifted distributed feedback resonators.}
\label{fig:scheme}
\end{figure}

We discuss a large-scale topological array laser composed of a number of site resonators. Figure \ref{fig:scheme}(a) shows a schematic implementation of such a laser based on Fabry-Pérot cavities, which can in principle be substituted by any other laser resonators. Each cavity supports a well defined single lateral mode and optically couples to neighboring cavities. Well-designed couplings between the cavities allow for the appearance of a topological lateral mode distributed over the nearly all of the cavity array, as we will describe shortly later. Electrodes are patterned on specific site cavities to selectively supply gain, which promotes lasing from the designed topological mode. We target a system including a few hundred resonators. If each resonator delivers $\sim$100 mW output, the topological laser could be operated as a 10W-class laser.

For theoretical analysis, we map this array laser to a simple tight binding model. We consider an array of single-mode resonators that resembles the Su-Schriffer-Heeger (SSH) model\cite{SSHcourse}. In the SSH model, the resonator chain is dimerized and its unit cell contains two resonators called A- and B-sites. When the coupling strengths for both the inter- and intra-unit cell hopping are the same, the model exhibits gapless energy bands in momentum space. Meanwhile, when the two coupling strengths are unequal, a gap appears between the two bands. For a SSH chain with a larger inter-cell coupling than the intra-cell coupling, its band topology becomes topologically non-trivial and topological localized modes emerge at the edges of the bulk chain according to the bulk-edge correspondence. 
More quantitatively, the topological properties of the energy bands can be characterized using Zak phases, which are defined by the integral of the Berry connection over the first Brillouin zone\cite{Zak1989}. For a topological band, its Zak phase takes a nonzero value and becomes $\pi$ when inversion symmetry is preserved in the system. 

To obtain the desired laser cavity mode, we interface two SSH chains at the center of the system, as schematically illustrated in Figure\ref{fig:scheme}(b). The two chains are topologically trivial and non-trivial, respectively. In this case, a single topological interface mode appears deterministically around the interface\cite{Henning2013,Ota2018}, with which we design a single mode laser. Since the other end of the topological SSH chain could support another edge state, we terminate the chain with an auxiliary site resonator strongly coupled to the bulk chain, by which we can suppress the emergence of the unwanted extra edge state.
Note that this configuration is similar to the design reported in Ref.\cite{Zhao2018}. However, they studied a tightly localized edge mode at the interface in a small lattice, in stark contrast with the current work investigating a broadly distributed interface mode in a large-scale lattice.
It is also interesting to note that the topological cavity structure illustrated in Figure 1(a) is reminiscent to that of distributed feedback lasers. Indeed, the laser mode in a lambda/4-shifted distributed feedback laser can be interpreted as a special case of a topological interface mode. Nevertheless, the discrete topological lattice model preserving chiral symmetry discussed in this paper will lead to a unique field distribution capable of robust single mode lasing, making a stark contrast with conventional distributed feedback lasers, as we will demonstrate later.

The system under consideration is described by the following Hamiltonian,  
\begin{eqnarray}
\begin{split}
\mathcal{H}
&=
\sum_{m=1}^{N+1}
\left(
i\gamma_{A,m} + \omega_{A,m}
\right)
\ket{m,A} \bra{m,A}
+
\sum_{m=1}^{N}
\left(
i\gamma_{B,m} + \omega_{B,m}
\right)
\ket{m,B} \bra{m,B} \\
&+
 \sum_{m=1}^{L} \left[\kappa_{2,m}
 \left(
\ket{m,B} \bra{m,A}+ h.c.
 \right)
+
\kappa_{1,m}
\left(
\ket{m+1,A} \bra{m,B}+ h.c.
 \right)
\right] \\
&+
\sum_{m=L+1}^{N} \left[\kappa_{1,m}
 \left(
\ket{m,B} \bra{m,A}+ h.c.
 \right)
+
\kappa_{2,m} 
\left(
\ket{m+1,A} \bra{m,B}+ h.c.
 \right)
\right],
\end{split}
\label{eq:Hamiltonian}
\end{eqnarray}
where $\omega_{A,m}$ and $\omega_{B,m}$ are the resonant frequencies of site A and B in a dimer $m$, respectively, while $\gamma_{A,m}$ and $\gamma_{B,m}$ denote gain and loss. 
Site-to-site coupling strengths are described by $ \kappa_{1,m}$ and $\kappa_{2,m}$.
We suppose $\kappa_{1,m} < \kappa_{2,m}$, such that the topological SSH chain always remains topological. 
The total number of the dimers and the number specifying the topological phase boundary are set as $N= 100$ and $L= 50$, respectively.
Thus, the number of sites in the trivial array becomes $n_{\rm tri}=2L$ = $100$, while that in the topological array does $n_{\rm topo}=2(N-L)+1$= $101$. 
The latter number includes the single auxiliary site at the end of the topological chain.
We neglect the presence of unwanted longitudinal modes in each resonator to simplify the analysis. This model is valid as long as the longitudinal modes behave independently or when assuming arrays of single longitudinal mode cavities such as $\lambda/4$-shifted distributed feedback resonators.
Note that, in this section, we consider an ideal case where we henceforth set ($\kappa_{1,m}$, $\kappa_{2,m}$, $\gamma_{A,m}$, $\gamma_{B,m}$)=($\kappa_{1}$, $\kappa_{2}$, $\gamma_{A}$, $\gamma_{B}$) and $\omega_{A,m} = \omega_{B,m} = \omega$ for any dimer $m$ , unless otherwise indicated.

\subsection{Eigenmodes in the absence of gain and loss}
\begin{figure}[h!]
\centering\includegraphics[width=14cm]{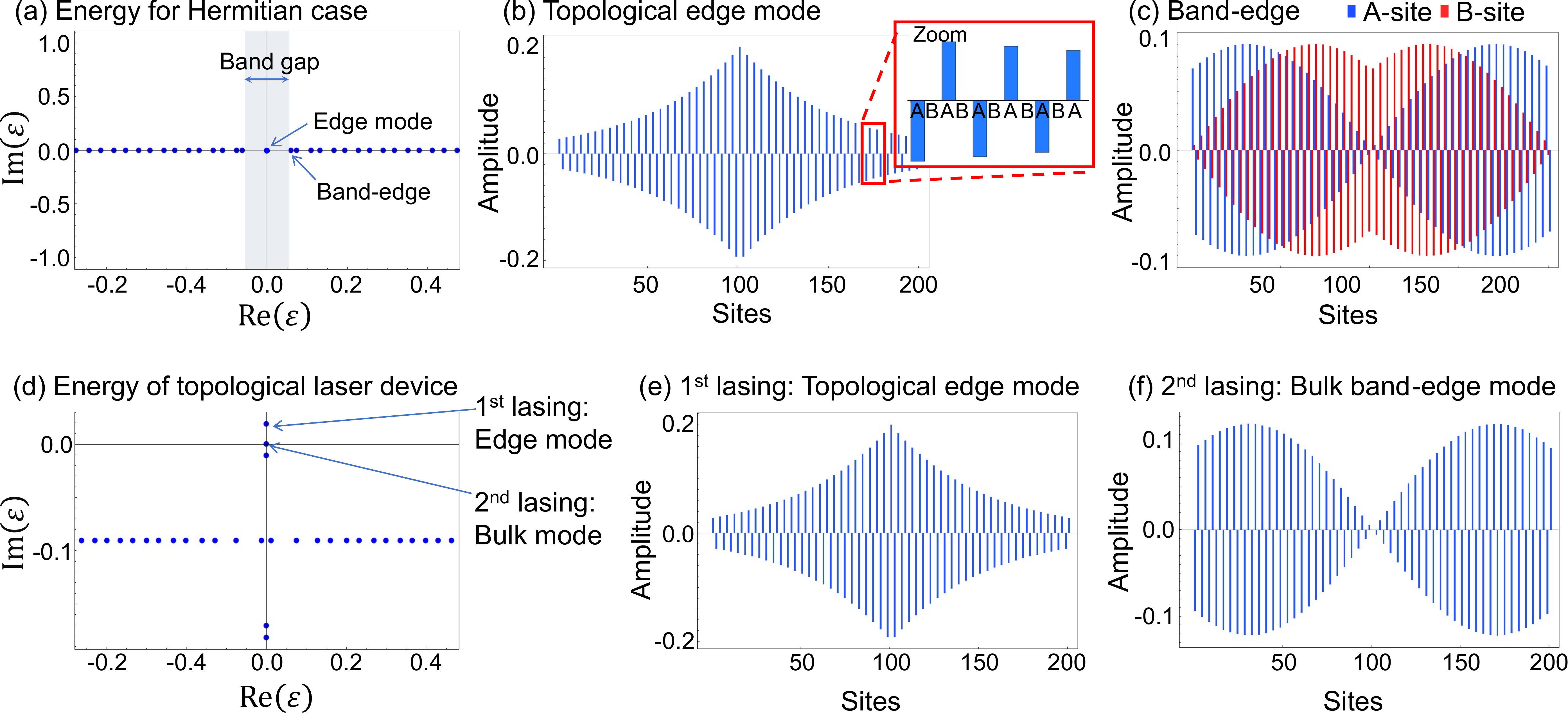}
\caption{(a) The eigenenergies $\varepsilon$ in the complex energy plane for the Hermitian case, zoomed around origin. (b) Spatial profile of the topological edge mode. The inset shows that the edge mode has non-zero amplitudes only on A-sites. 
(c) Spatial profile of the band-edge bulk mode for comparison. 
For (a)-(c), the parameters used are $\kappa_2/ \kappa_1= 1.04$, $\gamma_{\rm{loss}}=\gamma_{\rm{gain}}= 0$, $n_{\rm{tri}}= 100$ and $n_{\rm{topo}}= 101$.
(d) The eigenenergies $\varepsilon$ in the complex energy plane for the topological laser system with gain and loss, zoomed around origin.
(e) Spatial profile of the first lasing mode, i.e. topological edge mode. (f) Spatial profile of the second lasing mode steming from an amplified bulk mode, exhibiting non-zero amplitudes only on A-sites.
For (d)-(f), the parameters used are $\kappa_1$= 1.0, $\kappa_2$= 1.04, $\gamma_{\rm{loss}}$= 0.2, and $\gamma_{\rm{gain}}$= 0.219. The system size is $n_{\rm{tri}}= 100$ and $n_{\rm{topo}}= 101$.
In (b,c,e,f), blue and red bars indicate the amplitudes on A-site and B-site, respectively.
}
\label{fig:ideal}
\end{figure}

To understand the basic properties of the investigated system formulated in Eq.(\ref{eq:Hamiltonian}), we first analyze it in the absence of gain and loss. 
We set the coupling parameters to ($\kappa_{1,m}$, $\kappa_{2,m}$)= (1.0, 1.04), which serves as a basic parameter set for the subsequent discussion. We diagonalize the Hamiltonian and analyze the energy spectrum and the spatial profiles of the eigenmodes of the system. 
Figure \ref{fig:ideal}(a) shows computed eigenenergies $\varepsilon$ plotted in the complex energy plane. In the real energy spectrum, ${\rm Re}(\varepsilon)$, one can see an energy gap of approximately $2|\kappa_{1} - \kappa_{2}|$, 
in which an in-gap mode exists as expected from the topological design discussed above. 
The topological mode is fixed to the zero energy and the entire energy spectrum is symmetric with respect to the zero energy according to chiral symmetry existing in the system. 
Note that the chiral symmetry is preserved when Hamiltonian $H$ satisfies $\Gamma^{\dagger} H \Gamma$= -H with an operator $\Gamma^2=1$, where $\Gamma$ is Hermitian and unitary. In general, a lattice with chiral symmetry will be bipartite and has two sublattices such that no direct transition occurs between the same sublattice sites.
We inspect the spatial profile of the zero energy topological mode and plot this in Figure \ref{fig:ideal}(b). The mode profile distributes over the entire lattice with amplitudes only on A-sites\cite{Zhao2018,Parto2018}. The spatial profile is well described by an approximated analytical expression given as $a_m =(-\kappa_{1}/ \kappa_{2})^{|m-L|}\times a_L,  b_m=0$ for any $m$, where $a_m$ ($b_m$) is the field amplitude at $m$th A-site (B-site).  
The extent of the spatial profile depends on the ratio of coupling constants. The current ratio of $\kappa_2/\kappa_1$ = 1.04 is sufficiently small so that the edge mode profile is distributed over the entire 201 sites.
Figure \ref{fig:ideal}(c) shows a spatial profile of one of the two band-edge bulk modes. In contrast to the topological edge mode, the amplitudes are essentially equally distributed over both A and B-site. The difference of the mode profile suggests that lasing from the topological edge mode can be selectively promoted by supplying gain only to A-sites.

\subsection{Eigenmodes under the presence of gain and loss}

Next, we investigate the properties of the system when introducing gain and loss to assess the capability of single mode lasing.
To account for modal loss normally existing in photonic devices, we assume that all site resonators experience an uniform loss at a rate $\gamma_{\rm loss}$.
Then, we supply gain on the A-sites at a rate of $\gamma_{\rm gain}$.
Thus, we introduce $\gamma_A=\gamma_{\rm gain}-\gamma_{\rm loss}$ and $\gamma_B=-\gamma_{\rm loss}$ as imaginary onsite potentials across all the sites.
Figure \ref{fig:ideal}(d) shows representative eigenenergies in the complex energy plane with $\gamma_{\rm loss}= 0.2$ and $\gamma_{\rm gain}$= 0.218.
Most eigenstates show negative ${\rm Im}(\varepsilon)$ and are expected to behave as lossy states. In contrast, the topological edge state solely acquires an explicit positive ${\rm Im}(\varepsilon)$, indicating that the state becomes the first lasing mode in the system. This result confirms that our design can promote single mode lasing from the designed topological edge state broadly distributed in the lattice.
Meanwhile, in the plot, there is a bulk state with nearly zero real and imaginary energies, which, with additional gain, could be positive in the imaginary part and hence a second lasing state. 
The presence of such a bulk mode capable of lasing leads to the unwanted competition of lasing modes in the system. To stabilize the single mode lasing from the topological edge state, it is vital to design a system that suppresses lasing from the bulk modes.

The reason why the topological edge and bulk states under concern preferentially acquire non-negative ${\rm Im}(\varepsilon)$ can be understood from their mode profiles presented in Figure\ref{fig:ideal}(e) and (f). Both of their mode profiles have dominant amplitudes on A-sites, where gain is selectively supplied.  
We identified that the bulk mode with spatial profile only on the A-sites arises from a phase transition similar to that occurring in parity-time (PT) symmetric systems\cite{Nori2019,El-Ganainy2018,Bo2014}. 
Note that, while the gain and loss are totally balanced in PT symmetric systems, we consider a system with varied gain and fixed loss.
Subjected to a large supplied gain, some bulk modes experience a phase transition and choose to split in its imaginary energies (while in turn degenerate in real energies), which accompanies a drastic change of the field profiles. To gain more insight into the phase transition, we consider an infinite bulk SSH chain without any interface. In this case, the Hamiltonian represented in momentum space takes the form
\begin{eqnarray}
H(k)=
\begin{pmatrix}
i \gamma_A & \kappa_1+\kappa_2 {\rm e}^{-ika} \\
\kappa_1+\kappa_2 {\rm e}^{ika} & i \gamma_B \\
\end{pmatrix},
\end{eqnarray}
where $a$ is the lattice constant and $k$ is a wave number. For band edge modes supported at the Brillouin zone edge, the eigenvalues are given by 
$\varepsilon (\omega)
=\left[i(\gamma_A+\gamma_B) \pm \sqrt{-(\gamma_A-\gamma_B)^2 + 4(\kappa_1-\kappa_2) ^2}\right] /2$. 
The eigenvalues split in either real or imaginary part depending on the sign in the square root. One of the modes split in imaginary energy corresponding to the second lasing bulk mode in our case, as we will discuss later. Since we define 
$\gamma_A= \gamma_{\rm gain} -\gamma_{\rm loss}$ and 
$\gamma_B= -\gamma_{\rm loss}$, the critical gain that induces the imaginary energy splitting in the bulk mode is given by
\begin{eqnarray}
\gamma_{\rm gain}^{\rm{critical}} = 
2 \left| \kappa_1- \kappa_2\right|.
\label{eq:EP}
\end{eqnarray}
Across $\gamma_{\rm gain}^{\rm{critical}}$, one observes a phase transition in the bulk eigenstates. 
As has been anticipated from the Hamiltonian and the expression of eigenvalues above, the phase transition in the bulk system resembles that in PT symmetric systems. 
When $\gamma_{\rm gain} < \gamma_{\rm gain}^{\rm critical} $, the band-edge modes are in a phase analogous to the PT symmetric phase and exhibit mode profiles homogeneously-distributed for both sites. 
In contrast, when $\gamma_{\rm gain} > \gamma_{\rm gain}^{\rm critical}$, the band-edge modes are in a phase analogous 
to the broken-PT phase and therefore exhibit mode profiles that dominantly populate in either A- or B-site. 
The spatial profile of the bulk mode in Figure 2(f) shows the one in the broken phase. 
We note that $\gamma_{\rm gain}^{\rm critical}$ becomes larger when considering a bulk system with a finite size. For our system with 201 arrays, $\gamma_{\rm gain}^{\rm critical}$ is computed to be $ \sim 0.12$, instead of the analytical value  $\gamma_{\rm gain}^{\rm critical}$ = 0.08 for the infinite system with $|\kappa_1 - \kappa_2|$ = 0.04.

\subsection{Threshold gain difference}
\label{subsec:EP}

A way to assess the capability of single mode lasing is to measure the threshold gain difference among the lasing modes. 
In this work, the threshold gain for a mode is defined as the supplied gain at which the mode reaches Im$(\varepsilon)$=0. 
We will consider the threshold gain difference $\Delta\alpha$ between the first lasing topological mode and the second lasing bulk trivial mode. The former is defined to lase at $\gamma_{\rm{gain}}$=$\gamma_{\rm{th}}^{\rm{1st}}$ and the latter at $\gamma_{\rm{th}}^{\rm{2nd}}$, thus  $\Delta\alpha=\gamma_{\rm{th}}^{\rm{2nd}}-\gamma_{\rm{th}}^{\rm{1st}}$.
It is known that single mode lasing becomes more stable as $\Delta\alpha$ increases. 
The analysis based on $\Delta\alpha$ employs only the eigenmode analysis and thus is very simple, but nevertheless can effectively evaluate the single mode lasing stability. 

\begin{figure}[h!]
\centering\includegraphics[width=14cm]{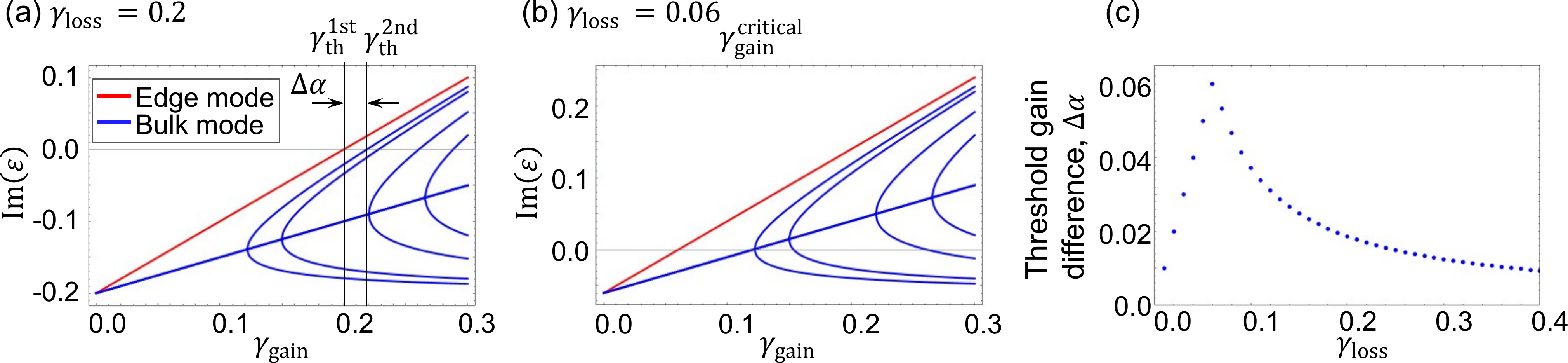}
\caption{(a)(b) Imaginary parts of eigenenergies of the system plotted as a function of supplied gain on A-site $\gamma_{\rm{gain}}$ for $\gamma_{\rm{loss}}$= 0.2 and 0.06, respectively. The red and blue lines indicate the energy of the edge mode and bulk modes.
(c) Loss dependence of the threshold gain difference $\Delta \alpha$. The parameters used are $\kappa_1$= 1.0 and $\kappa_2$= 1.04, with a finite system consisting of $n_{\rm tri}= 100$ trivial and  $n_{\rm topo}= 101$ topological cavities.}
\label{fig:net gain}
\end{figure}

Figure \ref{fig:net gain}(a) shows the calculated Im($\varepsilon$) as a function of $\gamma_{\rm{gain}}$ for a system with $\gamma_{\rm{loss}} = 0.2$. 
A loss value of $\gamma_{\rm{loss}}= 0.2$ is consistent with conventional Fabry Perot semiconductor lasers as we will discuss in section 4.
In the plot, it is clearly seen that the topological mode (colored in red) acquires gain much faster than the bulk modes (blue) and exhibits positive Im($\varepsilon$) at the lowest $\gamma_{\rm{gain}}$ among all the modes. 
The threshold gain for the edge mode $\gamma_{\rm{th}}^{\rm{1st}}$ equals to $\gamma_{\rm{loss}}$, since the edge mode distributes only on A-sites where gain is selectively supplied. With increasing $\gamma_{\rm{gain}}$, a bulk mode also reaches Im($\varepsilon$)=0 at $\gamma_{\rm{th}}^{\rm{2nd}} = 0.219$. Thus, $\Delta\alpha$ is equal to 0.019 in this particular example. 
If all bulk modes maintained an equal mode distribution on the A- and B-sites, $\Delta\alpha$ is expected to be $|\gamma_{\rm{loss}}|$ = 0.2, since they simply need additional gain to compensate the loss also in B-site. However, as already discussed above, some bulk modes undergo a phase transition that largely modifies their mode profiles. As such, the branched bulk modes acquire gain much faster than the rest of the bulk modes. This is the reason why $\Delta\alpha$ reduces in the case in Figure 3(a).
Meanwhile, the largest $\Delta\alpha$ can be obtained when the bulk mode reaches its lasing threshold $\gamma_{\rm th}^{\rm 2nd}$ at 
$\gamma_{\rm gain}^{\rm critical}$, that is $\gamma_{\rm th}^{\rm 2nd}$ =$\gamma_{\rm gain}^{\rm critical}$, which is more preferable for stable single mode lasing from the topological mode. This situation is realized in Figure 3(b), where $\gamma_{\rm loss}$ is set to 0.06. The overall behaviors of the Im$(\varepsilon)$ curves are exactly the same as those in Figure 3(a), except for the difference in the imaginary energy offset. This indicates that $\gamma_{\rm loss}$ is a critical factor for controlling $\Delta\alpha$. We note that $\gamma_{\rm loss}= 0.06$ may be too small to properly account for conventional loss in semiconductor lasers, which we will discuss in section 4.


In Figure 3(c), we evaluate $\Delta\alpha$ as a function of $\gamma_{\rm{loss}}$ for the system defined in Figure 1(b) with $\kappa_1$= 1.0 and $\kappa_2$= 1.04. The plot of $\Delta\alpha$ shows a peak at $\gamma_{\rm loss}$ = 0.06 where $\gamma_{\rm th}^{\rm 2nd}=\gamma_{\rm gain}^{\rm critical}$ holds, as discussed in Figure 3(b). 
For the region of $\gamma_{\rm{loss}} < 0.06$, there is a linear increase of $\Delta\alpha$ with increasing $\gamma_{\rm{loss}}$. In this situation, $\gamma_{\rm th}^{\rm 2nd}$ is lower than $\gamma_{\rm gain}^{\rm critical}$ and the second lasing starts before the bulk modes get branched.  
For the region of $\gamma_{\rm loss} > 0.06$, there is a monotonic decrease of $\Delta\alpha$ with increasing $\gamma_{\rm{loss}}$. In this situation, the second bulk-mode lasing occurs from a branched mode and thus $\gamma_{\rm{th}}^{\rm{2nd}}$ becomes closer to $\gamma_{\rm{th}}^{\rm{1st}}$.

We here summarize the points to be considered for increasing $\Delta\alpha$ in our system. (i) The maximum possible $\Delta\alpha$ is obtained when $\gamma_{\rm th}^{\rm 2nd}$ = 
$\gamma_{\rm gain}^{\rm critical}$. 
(ii) A large $\gamma_{\rm gain}^{\rm critical}$ is preferable for enhancing $\Delta\alpha$. 
(iii) $\gamma_{\rm gain}^{\rm critical}$ can be increase by increasing $|\kappa_1 - \kappa_2|$, while $|\kappa_1/\kappa_2|$ should be close to one for maintaining a large field extent of the topological mode. 
Thus, one should take large $\kappa_1$ and $\kappa_2$ with $|\kappa_1/\kappa_2| \sim 1$. 
(iv) There is an optimal $\gamma_{\rm loss}$ in the system with respect to $\gamma_{\rm gain}^{\rm critical}$ for maximizing $\Delta\alpha$. For semiconductor array lasers based on conventional lossy resonators, the above discussion suggests that it is important to employ low-loss resonators with high resonator-resonator couplings. We will revisit more practical considerations for achieving a single-mode large-area topological laser in section 4.

\section{Effects of disorders and long-range interactions on the single-mode lasing operation}

In this section, we evaluate the stability of the single mode lasing under the presence of imperfections by primarily considering $\Delta\alpha$.
We examine the effects of inhomogeneous coupling strengths and resonator frequencies that are the most likely types of disorder induced by fabrication imperfections. 
Previous works have studied the effect of such disorders in 1D SSH models, however, most of them are focusing on the properties of tightly localized topological edge modes\cite{Amo2017,Prodan2014,Platero2019,Bauer2019,Kennett2020}. 
In contrast, our interest lies in the broadly distributed topological edge mode and its stability of single mode lasing in competition with a bulk mode.
We also discuss the effect of interactions between next-nearest neighbor resonators, which are likely to occur in optical resonator arrays in the course of increasing nearest neighbor couplings.


\subsection{Inhomogeneous coupling strengths}

First, we investigate the effects of inhomogeneity in the coupling strengths on the laser array systems discussed so far, i.e. those constructed with $\kappa_{1}$= 1.0 and $\kappa_{2}$= 1.04 for 201 sites.
We prepare coupling strengths randomly distributed among all sites by generating different sets of Gaussian random variables with means $\kappa_{1}$= 1.0 (for intra-dimer coupling) and $\kappa_{2}$= 1.04 (inter-dimer) and a common standard deviation of randomness $r_{\kappa}$. For each set of parameters with the randomness, we solve the Hamiltonian in Eq.(\ref{eq:Hamiltonian}) by diagonalization.
In order to study $r_{\kappa}$ dependence of the laser system, we generate 100 different sets of parameters for each $r_{\kappa}$, and average the outcomes.
The error bar represents the half of the standard deviation $\sigma/2$, throughout this section.
We note that the disorder discussed here can be interpreted as random distances between the site resonators, hence it only breaks parity symmetry, while preserving chiral symmetry.

\begin{figure}[h!]
\centering\includegraphics[width=13cm]{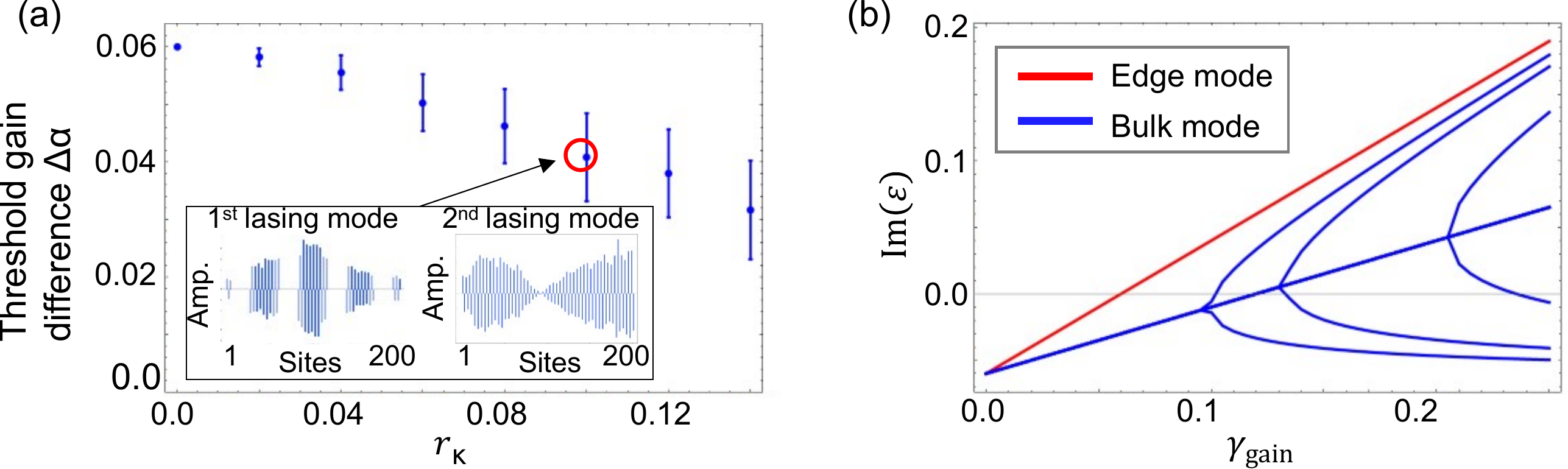}
\caption{(a) Threshold gain difference between the first and second lasing modes as a function of coupling disorder $r_{\kappa}$ in
a finite system consisting of $n_{\rm{tri}}= 100$ trivial and $n_{\rm{topo}}=101$ topological cavities. The insets show representative mode profiles of the edge mode and bulk mode. Blue bars indicate the amplitudes on A-site.
(b) Imaginary parts of eigenenergies of the system plotted as a function of supplied gain on A-site $\gamma_{\rm gain}$. The red and blue lines indicate the energy of the edge mode and bulk modes, respectively.
The parameters used are $\kappa_{1}$= 1.0, $\kappa_{2}$= 1.04 and $\gamma_{\rm loss}$= 0.06. In (b), the randomness is $r_{\kappa}$= 0.1.}
\label{fig:inhomo coupling}
\end{figure}

Figure \ref{fig:inhomo coupling}(a) shows the computed threshold gain differences $\Delta\alpha$'s for a system subject to $\gamma_{\rm{loss}} = 0.06$. This is the case for realizing the largest $\Delta\alpha$ for the disorder-free case. As the randomness or $r_{\kappa}$ increases, a decreased $\Delta\alpha$ is observed. However, $\Delta\alpha$ remains $\sim$70$\%$ of the maximum even when $r_{\kappa}$ = 0.1, where the strength of randomness as the standard deviation exceeds the bandgap of the infinite Hermitian system, $2|\kappa_1-\kappa_2|$ = 0.08. This result indicates the robustness of the single mode lasing from the resonator array device. 
In the current case, the threshold gain for the first lasing mode, $\gamma_{\rm th}^{\rm 1st}$, remains unchanged even when introducing the disorders. This is a consequence of the preserved chiral symmetry, which leads to a zero energy mode with its mode amplitude only on A-sites, thus always reaching the threshold gain exactly when compensating the loss in A-sites. Therefore, the observed decrease of $\Delta\alpha$ arises solely from the decrease of the threshold gain for the second lasing mode $\gamma_{\rm th}^{\rm 2nd}$. 
As discussed in the previous sections, $\gamma_{\rm th}^{\rm 2nd}$ diminishes for a lower $\gamma_{\rm gain}^{\rm critical}$, which scales with $2|\kappa_1-\kappa_2|$ for the unperturbed case. We consider that the introduction of randomness masks the difference between the couplings by $\kappa_1$ and $\kappa_2$ and hence effectively reduces $|\kappa_1-\kappa_2|$. 
Accordingly, we found a gradual reduction of the width of average bandgap in the system with increasing $r_{\kappa}$.
To further verify the above discussion, we computed the spatial profile of the first and second lasing mode for the case with $r_{\kappa}$ = 0.1, as plotted in the insets in Fig. 4(a). We plot typical mode profile providing the average $\Delta\alpha$ among the 100 trials. The mode profiles resemble those computed for $r_{\kappa}$ = 0. This observation confirms that the first lasing mode originates from the topological interface mode and the second one originates from the bulk edge mode as observed in the unperturbed case. 
Figure 4(b) shows a computed ${\rm Im}(\varepsilon)$ as a function of $\gamma_{\rm gain}$ for the parameter set used in the insets in Fig. 4 (a). As anticipated above, one can see the reduction of $\gamma_{\rm gain}^{\rm critical}$ to 0.10 and hence of $\Delta\alpha$ by 70$\%$ in comparison to the disorder-free case in Fig. 3(b). 
Overall, it was found that the topological mode robustly behaves even under the presence of the disorder for coupling strengths with $r_{\kappa} > 2|\kappa_1 - \kappa_2|$. 
In section 4, we will quantitatively discuss $r_{\kappa}$ by referring a required accuracy in the actual device fabrication for an example case.
It is interesting to note that the interface of the two topologically distinct chains may effectively remain even with such a large $r_{\kappa}$, as indicated in the spatial profile of the zero energy mode plotted in the insets in Fig. 4(a). The mode has a peak near the center of the system, where the interface is originally located. 
Another important note is that a very similar tendency was observed for the case that only replaces $\gamma_{\rm{loss}}$ from 0.06 to 0.2. Even in this case, we observed a reduction of the average $\Delta\alpha$ by 70$\%$ when $r_{\kappa}$ = 0.1. This result implies that loss does not essentially alter the behavior of the system subject to inhomogeneous coupling strengths.

\subsection{Inhomogeneous site resonator frequencies}

Next, we perform calculations for the cases with fluctuations in the resonance frequencies of the site resonators. We treat inhomogeneity in the resonator detunings $\Delta$ after subtracting a common frequency offset $\omega$ from the Hamiltonian in Eq. (1). For the perfectly regular case, $\Delta$ equals to zero for any $m$th resonator. 
We prepare 100 sets of random $\Delta$'s distributed by Gaussian random variables with means $\Delta = 0$ and the standard deviation $r_{\Delta}$.
We introduce each set of generated random detunings in Eq.(\ref{eq:Hamiltonian}) and solve it by diagonalization for the system with $\kappa_1$ = 1.0 and $\kappa_2$ = 1.04. The ways of averaging the data for each $r_{\Delta}$ and of its plot are the same as in the previous section.

\begin{figure}[h!]
\centering\includegraphics[width=13cm]{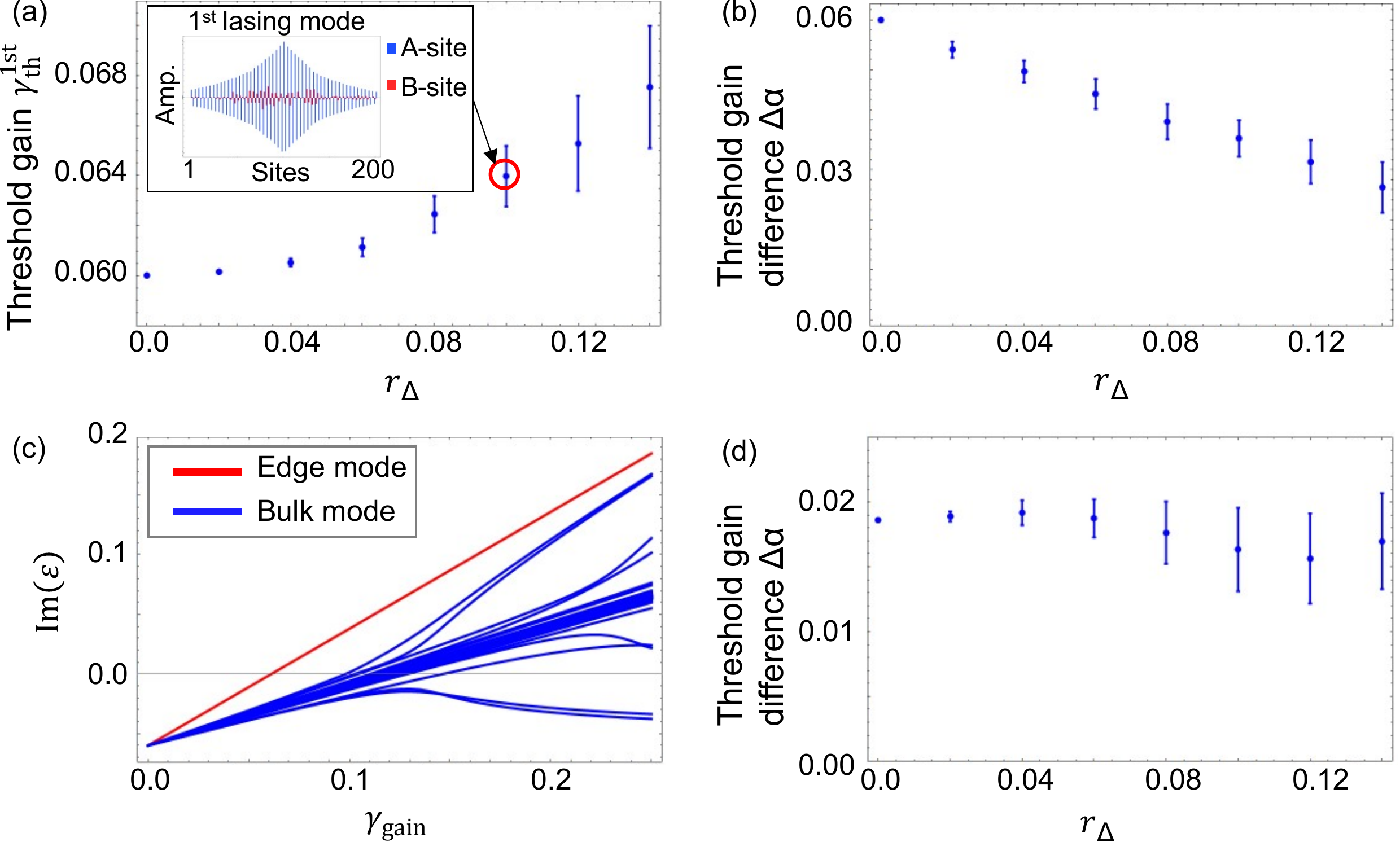}
\caption{
(a) Threshold gain of the first lasing mode as a function of strength of inhomogeneity in detuning $r_{\Delta}$ in
a finite system consisting of $n_{\rm{tri}}= 100$ trivial and $n_{\rm{topo}}= 101$ topological cavities. The inset shows a representative sample of the edge mode profile for $r_{\Delta}= 0.1$ where blue and red bars indicate the amplitudes on A-site and B-site, respectively.
(b) Threshold gain difference $\Delta\alpha$.
(c) Imaginary parts of eigenenergies versus supplied gain on A-site $\gamma_{\rm gain}$. The red and blue lines indicate the energy of the edge mode and bulk modes, respectively.
(d) Threshold gain difference $\Delta\alpha$ for the system with higher resonator loss of $\gamma_{\rm loss}$= 0.20.
The coupling constant is $\kappa_2/\kappa_1$= 1.04 and the loss is $\gamma_{\rm loss}$= 0.06 in (a)-(c) and $\gamma_{\rm loss}$= 0.20 in (d).
}
\label{fig:inhomo detuning}
\end{figure}

Figure \ref{fig:inhomo detuning}(a) shows the average $\gamma_{\rm{th}}^{\rm{1st}}$ with varying $r_{\Delta}$ for a system with $\gamma_{\rm{loss}}$ = 0.06. Unlike the case with the coupling disorders, the average $\gamma_{\rm{th}}^{\rm{1st}}$ slightly increases with $r_{\Delta}$. In the system with non-zero $r_{\Delta}$, chiral symmetry is broken and thus the topological mode acquires a field amplitude also in lossy B-sites, resulting in the increase of $\gamma_{\rm th}^{\rm 1st}$. This behavior can be confirmed in the mode profile in the inset in Figure \ref{fig:inhomo detuning}(a) calculated for a representative example when $r_{\Delta}$ = 0.1. The mode profile consists mainly of the original topological interface mode, but slightly contains B-site amplitudes, which is consistent with the modest increase of $\gamma_{\rm th}^{\rm 1st}$.
Figure \ref{fig:inhomo detuning}(b) shows average $\Delta\alpha$ calculated for the system with $\gamma_{\rm loss}$ = 0.06. A monotonic decrease of $\Delta\alpha$ is found, which is much larger amount than the increase in $\gamma_{\rm{th}}^{\rm 1st}$. Thus, the drop of $\Delta\alpha$ is expected to stem from a decrease of $\gamma_{\rm th}^{\rm 2nd}$. Figure \ref{fig:inhomo detuning}(c) shows the computed ${\rm Im}(\varepsilon)$ for the system discussed in the inset in Figure \ref{fig:inhomo detuning}(a). As anticipated, an earlier growth of ${\rm Im}(\varepsilon)$ for a bulk mode is seen when increasing $\gamma_{\rm gain}$, making the $\Delta\alpha$ smaller. 
In the plot, it is seen that the phase transition in the bulk modes is blurred and a diagonal bundle of the bulk modes are formed. These are the consequences of the symmetry breaking by the fluctuating $\Delta$. While sharp branches of the bulk modes are not observed in Figure \ref{fig:inhomo detuning}(c), the overall behaviors of branched curves in ${\rm Im}(\varepsilon)$ are similar with those in Figure 3(b), in particular for large $\gamma_{\rm gain}$ roughly over 0.15. This comparison suggests that the fluctuation in $\Delta$ mainly influences how the ${\rm Im}(\varepsilon)$ curves branch out from the bulk mode bundle.
Figure \ref{fig:inhomo detuning}(d) shows $\Delta\alpha$ computed for the system with $\gamma_{\rm loss}$ = 0.2. In contrast to the case with lower loss, the computed $\Delta\alpha$s are less sensitive with large $\gamma_{\rm loss}$. This is because introducing the fluctuating $\Delta$ does not alter the overall behaviors of ${\rm Im}(\varepsilon)$ curves in particular for large $\gamma_{\rm gain}$, at which $\Delta\alpha$ is measured for the case of $\gamma_{\rm loss}$ = 0.2. In other words, for large $\gamma_{\rm gain}$, the relationship between the ${\rm Im}(\varepsilon)$ curves of the topological mode and the competing bulk mode does not change largely, neither does $\Delta\alpha$.


\subsection{Next-nearest-neighbor cavity coupling}

The discussion in section \ref{sec:ideal} reveals that larger coupling strengths between the site resonators are advantageous for achieving a large $\Delta\alpha$ and thus for stable single mode lasing from a broadly-distributed topological edge mode.
Cavity array designs for increasing the coupling strengths between the nearest neighbor (NN) cavities may inevitably induce non-negligible next-nearest-neighbor (NNN) couplings, which will break chiral symmetry and thus could modify the performance of the laser device.
In this section, we analyze the influence of NNN couplings on the investigated array laser. 

\begin{figure}[h!]
\centering\includegraphics[width=13cm]{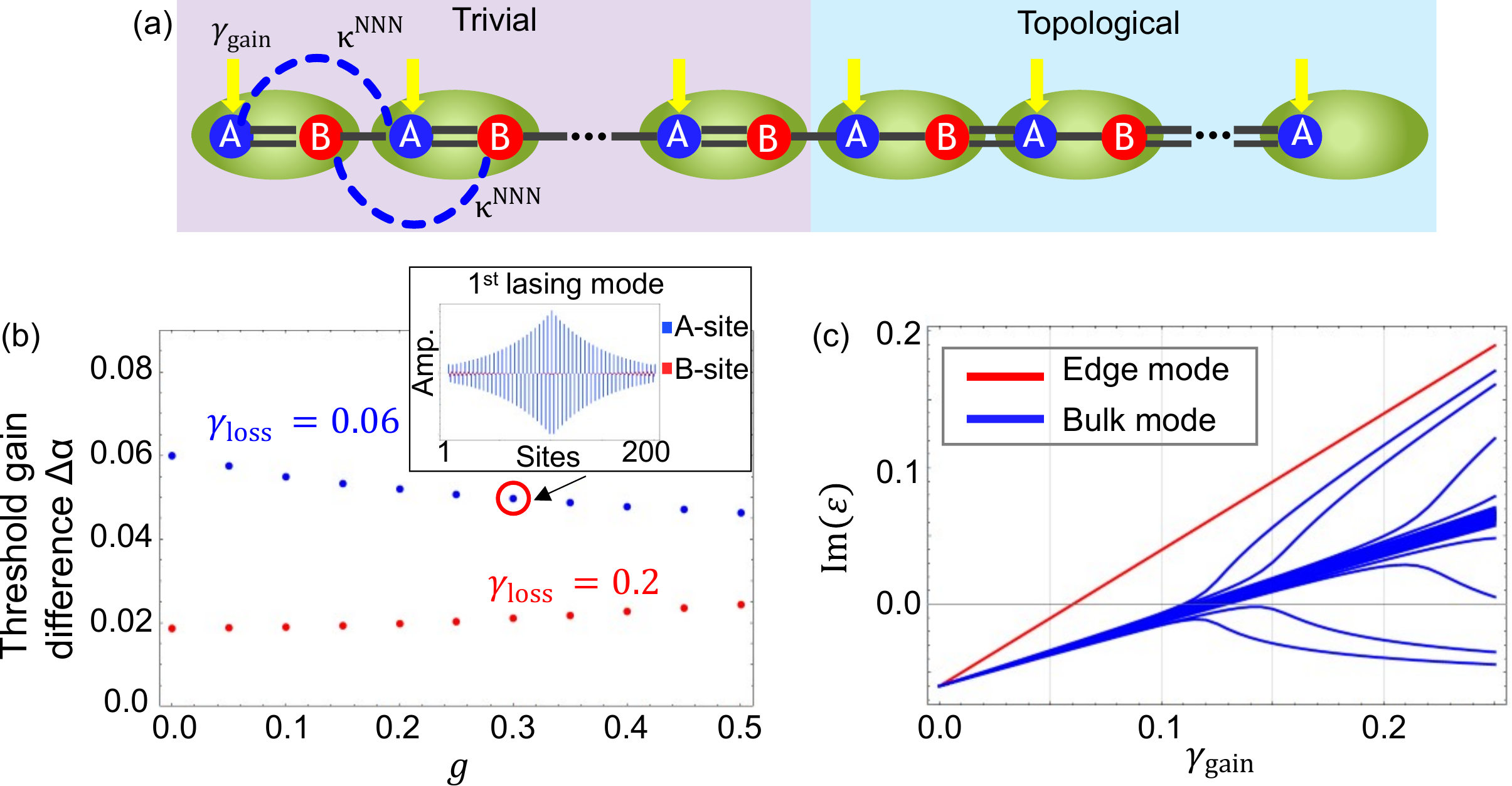}
\caption{
(a) Extended tight-binding model for the topological laser, including the next-nearest-neighbor (NNN) couplings. Nearest-neighbor (NN) couplings and NNN coupligs are given as {$\kappa_1^{\rm{NN}},\kappa_2^{\rm{NN}}$} and $\kappa^{\rm{NNN}}$, respectively. All sites are subject to loss at a rate of $\gamma_{\rm{loss}}$, while gain $\gamma_{\rm{gain}}$ is additionally supplied only to the A-sites.
(b) Threshold gain difference $\Delta\alpha$ as a function of ratio $g$ of NNN couplings to NN couplings in a finite system consisting of $n_{\rm{tri}}= 100$ trivial and $n_{\rm{topo}}= 101$ topological cavities. 
Blue and red dots are for the loss $\gamma_{\rm loss}$= 0.06 and 0.2, respectively.
The inset in (b) provides a representative sample of the edge mode profile for $g= 0.3$ where blue and red bars indicate the amplitudes on A-site and B-site, respectively.
(c) Imaginary parts of eigenenergies versus supplied gain on A-site $\gamma_{\rm gain}$. The red and blue lines indicate the energy of the edge mode and bulk modes, respectively.
The parameters used are $\kappa_1$= 1.0, $\kappa_2$= 1.04 and the loss is set to $\gamma_{\rm{loss}}$= 0.06 in (c).}
\label{fig:NNN}
\end{figure}

Figure \ref{fig:NNN}(a) explains the model we consider in this section. 
We define the ratio of the NNN couplings to NN couplings by a factor $g$: $g=\kappa^{\rm{NNN}}/\kappa_1^{\rm{NN}}$, where $\kappa^{\rm{NN}}$ and $\kappa^{\rm{NNN}}$ denote the coupling strength between NN and NNN cavities, respectively.
We add a term of the NNN couplings to the Hamiltonian in Eq. (1) with $\kappa_1$ = 1.0 and $\kappa_2$ = 1.04 and solve it by diagonalization.
Figure \ref{fig:NNN}(b) shows computed $\Delta\alpha$ as a function of $g$. The plot contains two curves calculated for the system with $\gamma_{\rm{loss}}$ = 0.06 and 0.2, respectively. Interestingly, both two curves do not show significant changes in $\Delta\alpha$ even when increasing the strength of NNN coupling as far as $g < 0.5$. For both cases, the change in $\Delta\alpha$ is only 20$\%$ at maximum. These behaviors can be understood by the combination of the computed mode profile and ${\rm Im}(\varepsilon)$, as plotted in Figure 6(b) and (c) for the case with $\gamma_{\rm{loss}}$ = 0.06. We find that the introduction of the NNN coupling do not largely modify the mode profile and the ${\rm Im}(\varepsilon)$ curves compared to those computed with only NN coupling. We note that, under the presence of the NNN couplings, the topological edge mode includes B-site amplitudes in its mode profile as shown in the inset in Figure 6(b) and the bulk modes resolve their degeneracy and form a bundle in ${\rm Im}(\varepsilon)$ curves as in Figure 6(c). These are the results of the absence of chiral symmetry in the system. We also note that $g > 0.5$ may be unlikely to occur for laser arrays based on evanescent mode coupling. Since evanescence fields exponentially decay in space, NN coupling is tend to be much larger than NNN coupling for most laser cavities. These insights obtained in this section are encouraging for increasing $\Delta\alpha$ by strengthening NN coupling with virtually ignoring the increase of NNN coupling.



\section{Discussion}


In this section, we discuss practically-achievable $\Delta\alpha$ for the topological array laser system that we discussed in the previous sections. The device under consideration consists of 201 site resonators with $\kappa_1$ = 1.0 and $\kappa_2$ = 1.04 so that it supports a broadly-distributed single topological edge mode. 
First, we estimate achievable strengths of $\kappa_1$ and $\kappa_2$ for conventional ridge-waveguide Fabry-Perot cavities based on GaAs/AlGaAs materials as an example. By choosing the ridge width of 1.4 {\textmu}m, height of 1.6 {\textmu}m and the gap between the ridges of 0.5 {\textmu}m, the coupling strengths of $\sim$100 cm$^{-1}$ is found to be possible by simulations using a finite element method. 
Thus, in the following discussion, we mainly consider the cases with $\kappa_1$ = 100 cm$^{-1}$ and $\kappa_2$ = 104 cm$^{-1}$. Note that, the fluctuation of $\kappa_1$ by 10$\%$ (corresponding to the case with $r_{\kappa} \sim$ 0.1) can only happen when the ridge-to-ridge distance varies more than 150 nm. This level of fabrication imperfection is unlikely to occur using standard semiconductor processing technologies.

Once fixing the coupling strengths, the most critical factor determining $\Delta\alpha$ is the resonator loss. 
From Figure 3(c), it is possible to deduce a $\Delta\alpha$ of 0.019 for a loss of $\gamma_{\rm loss}$ = 0.2. 
This case corresponds to $\Delta\alpha$ of 1.9 cm$^{-1}$ when $\kappa_1$ = 100 cm$^{-1}$ and thus $\gamma_{\rm loss}$ = 20 cm$^{-1}$ (Table 1), which is a moderate loss for typical semiconductor lasers with careful design and fabrication.
Given the previously reported values for semiconductor lasers\cite{Noda2019}, $\Delta\alpha$ of 1.9 cm$^{-1}$ could lead to stable single mode lasing in the device.
As indicated in Figure 3(c), the maximum possible $\Delta\alpha$ can be obtained at the optimal point of the loss setting with $\gamma_{\rm loss}$ = 0.06. 
For a system with $\kappa_1$ = 100 cm$^{-1}$, these values are converted into $\Delta\alpha$ = 6 cm$^{-1}$ and $\gamma_{\rm loss}$ = 6 cm$^{-1}$.
While $\Delta\alpha$ of 6 cm$^{-1}$ may be regarded as a sufficiently high for stable single mode lasing, the loss of $\gamma_{\rm{loss}}$ = 6 cm$^{-1}$ is too low when assuming the use of standard semiconductor lasers. 
In general, the optical loss in a semiconductor Fabry-Perot laser with zero carrier injection is composed of optical propagation loss, mirror loss and absorption in the active material. For a GaAs/AlGaAs ridge waveguide, the propagation loss can be reduced to about a few cm$^{-1}$, while mirror loss becomes 6 cm$^{-1}$ even for a 2 mm long cavity with a high reflection coating at a facet. Therefore, if including photon absorption in the unpumped active material, it is rather hard to realize the resonator optical loss of $\gamma_{\rm{loss}}$ = 6 cm$^{-1}$ to achieve the maximum possible $\Delta\alpha$ = 6 cm$^{-1}$. 
\begin{table}[htb]
\begin{center}
\caption{Values of $\Delta\alpha$ and their corresponding $\gamma_{\rm loss}$ for two representative coupling strength $\kappa_1$.}
  \begin{tabular}{l||c|c|c} \hline
     & Maximum $\Delta\alpha$  & $\gamma_{\rm loss}$ at maximum $\Delta\alpha$ & $\Delta\alpha$ at $\gamma_{\rm loss}$ = 20cm$^{-1}$  \\  \hline
    $\kappa_1$ = 100 cm$^{-1}$ & 6 cm$^{-1}$ & 6 cm$^{-1}$ & 1.9 cm$^{-1}$ \\ \hline
    $\kappa_1$ = 150 cm$^{-1}$ & 9 cm$^{-1}$ & 9 cm$^{-1}$ & 4.2 cm$^{-1}$ \\ \hline
  \end{tabular}
  \end{center}
\end{table}

There are several possible ways to significantly reduce material absorption loss in semiconductor laser resonators for achieving large $\Delta\alpha$. One straightforward way is to electrically pump lossy resonators. By introducing an additional gain of $\gamma^{\rm B}_{\rm gain}$ to B-sites, the loss effectively reduces and thus $\gamma_{\rm gain}^{\rm critical}$ increases by $\gamma^{\rm B}_{\rm gain}$: i.e. Eq.(\ref{eq:EP}) is modified to $\gamma_{\rm gain}^{\rm critical} = \pm 2 \left| \kappa_1- \kappa_2 \right| +\gamma^{\rm B}_{\rm gain}$. 
By recalling the fact that the largest $\Delta\alpha$ can be realized when $\gamma_{\rm th}^{\rm 2nd}$ =$\gamma_{\rm gain}^{\rm critical}$ as shown in Figure 3(b), this configuration may bring a powerful solution to reach stable single-mode lasing for a system with large $\gamma_{\rm{loss}}$.
When $\gamma_{\rm loss}$ = 20 cm$^{-1}$, $\Delta\alpha$ can take the maximum possible value of 6 cm$^{-1}$ by injecting $\gamma_{\rm gain}^{\rm B}$ of 14 cm$^{-1}$.
Another possibility for reducing $\gamma_{\rm{loss}}$ is to use tailored gain materials and structures. It has been predicted that sufficiently p-doped semiconductor quantum dots can quench inter-band light absorption while maintaining high differential gain under electrical current injection\cite{Arakawa1982}. 
Thus, $\gamma_{\rm{loss}}$ will be reduced for both A- and B-sites. However, the suppression of free-carrier absorption induced by the p-doping could be another experimental issue for achieving a low $\gamma_{\rm{loss}}$.
The use of buried heterostructures \cite{NTTburied2016} could also be used to selectively reduce $\gamma_{\rm{loss}}$ from B-sites by eliminating active materials only from B-sites. 
Using the above-mentioned means, the absorption in the active materials may be suppressed so that the optical loss of $\gamma_{\rm{loss}}$ = 6 cm$^{-1}$ can be achieved which is the optimal for ensuring large $\Delta\alpha$.

It is also interesting to discuss other ways to improve $\Delta\alpha$ for large $\gamma_{\rm{loss}}$. 
As we have already observed, the introduction of NNN couplings is not largely detrimental to the single mode operation. 
Therefore, $\Delta\alpha$ can be enlarged by increasing NN couplings $\kappa_1$ as shown in Table 1.
When $\kappa_1$=150 cm$^{-1}$, $\Delta\alpha$ will be increased to 4.2 cm$^{-1}$ even in the case of $\gamma_{\rm loss}$ = 20 cm$^{-1}$.
Note that the increased NN couplings also relaxes the condition for achieving the maximum $\Delta\alpha$.
In such cases with large NN couplings, NNN and very long range couplings will become significant to determine the band structures, making the system more similar to photonic crystals where the long-range interactions are dominant. 
Designs of topological edge mode lasers using such structures toward high power output will be an interesting topic of further research. 

Another interesting approach for increasing $\Delta\alpha$ is to use additional auxiliary lossy resonators. According to Figure 2(e) and (f), the mode profiles of the topological edge mode and the competing bulk mode differ largely in term of their envelope: the bulk mode shows a greater extent to the exterior of the system. Therefore, it could be possible to selectively load more loss on the bulk mode by terminating the system with auxiliary loss sites. 
We examined this idea for the system with $\kappa_1$= 100 cm$^{-1}$ by adding 10 lossy resonators with the same amount of loss for each termination, $\gamma_{\rm loss}$ =20cm$^{-1}$. We observed an increase of $\Delta\alpha$ from 1.9 cm$^{-1}$ to 2.4 cm$^{-1}$ in this case. Note that this approach does not work well for the cases with low $\gamma_{\rm loss}$ less than 6 cm$^{-1}$. In such cases, the competing bulk mode lases before $\gamma_{\rm gain}$ reaching $\gamma_{\rm gain}^{\rm critical}$ and its mode profile differs from that in Figure 2(f), leading to a small overlap with the additional lossy sites. 
Before closing this section, we briefly address another important factor that dictates the capability of the topological single mode laser, namely the presence of damage threshold. The laser mode profile presented in Figure 2(e) shows a peak at the center, at which photon density will be larger than the rest of the resonator sites. The laser will be operated so as not to exceed the damage threshold at the center resonator, suggesting that the rest of the resonator sites cannot deliver their maximum output power. This will clearly reduce the maximum possible output power from the system. Topological resonator designs that support flat-top mode shapes is one solution to this issue. Such designs are available by tailoring the distribution of the coupling strengths among the resonators. We will report the impact of such designs on the laser performance elsewhere.

\section{Summary}

We investigated a fundamental model of broadly-distributed single-mode topological edge mode laser in the tight-binding approximation.
We considered a sizable system consisted of 201 site resonators that potentially lead to a 10W-class laser by assuming that each resonator delivers $\sim$100 mW output power. 
We clarified the conditions for single-mode operation by calculating threshold gain differences $\Delta\alpha$ between the first lasing edge-mode and the second lasing bulk mode as an important factor for evaluating the stability of the single-mode operation.
Below is a summary of what we found through the discussion:
(a) Under ideal conditions, $\Delta\alpha$ depends on the coupling strengths $\kappa_{1}$, $\kappa_{2}$ and the loss $\gamma_{\rm loss}$. There exists an optimal loss for each combination of the coupling strengths. For a system based on semiconductor lasers, large $\kappa_{1}$ and $\kappa_{2}$ with $|\kappa_1/\kappa_2|\sim 1$ and small $\gamma_{\rm loss}$ are most preferable for stable single mode lasing.
(b) The single-mode operation of the edge mode is robust against disorders in coupling strengths and resonator detunings.
(c) The topological laser is insensitive to the addition of resonator couplings among NNN sites. This suggests that one can design laser systems with large $\kappa_{1}$ and $\kappa_{2}$ while virtually ignoring the influence of the NNN couplings.
(d) When assuming a set of realistic parameters for semiconductor lasers, $\Delta\alpha$ reaches a few cm$^{-1}$, which could be large enough for stable single-mode lasing.
To conclude, we provided significant insights for topological lasers in the context of realizing high power lasers.
This work may open up a new pathway for practical applications of topological photonics.

\section*{Acknowledgements}
Authors thank JSPS KAKENHI Grant Number JP21J40088, MEXT KAKENHI Grant Number JP15H05700, JP15H05868 and 17H06138, JST CREST (JPMJCR19T1) and NEDO.
T.B. is supported by the National Natural Science Foundation of China (62071301); State Council of the People’s Republic of China (D1210036A); NSFC Research Fund for International Young Scientists (11850410426); NYU-ECNU Institute of Physics at NYU Shanghai; the Science and Technology Commission of Shanghai Municipality (19XD1423000); the China Science and Technology Exchange Center (NGA-16-004).

\bibliographystyle{unsrt} 
\bibliography{main}

\end{document}